\documentclass
[superscriptaddress,secnumarabic,amssymb,amsmath,nobibnotes,aps,prd,showkeys,showpacs,nofootinbib]{revtex4}%
\usepackage{graphicx}
\usepackage{epsf}
\usepackage{bm}
\usepackage{amsmath}
\usepackage{amsfonts}
\usepackage{amssymb}
\usepackage{epstopdf}%
\providecommand{\U}[1]{\protect\rule{.1in}{.1in}}
\newcommand{\be}{\begin{equation}}
\newcommand{\ee}{\end{equation}}

\newcommand{\mincir}{\raise
-3.truept\hbox{\rlap{\hbox{$\sim$}}\raise4.truept\hbox{$<$}\ }}
\newcommand{\magcir}{\raise
-3.truept\hbox{\rlap{\hbox{$\sim$}}\raise4.truept\hbox{$>$}\ }}

\begin{document}
\title{Classical and Quantum Solutions in Brans-Dicke Cosmology with a Perfect Fluid}
\author{Andronikos Paliathanasis}
\email{anpaliat@phys.uoa.gr}
\affiliation{Instituto de Ciencias F\'{\i}sicas y Matem\'{a}ticas, Universidad Austral de
Chile, Valdivia, Chile}
\author{Michael Tsamparlis}
\email{mtsampa@phys.uoa.gr}
\affiliation{Faculty of Physics, Department of Astrophysics - Astronomy - Mechanics,
University of Athens, Panepistemiopolis, Athens 157 83, Greece}
\author{Spyros Basilakos}
\email{svasil@academyofathens.gr}
\affiliation{Academy of Athens, Research Center for Astronomy and Applied Mathematics,
Soranou Efesiou 4, 11527, Athens, Greece}
\author{John D. Barrow}
\email{jdb34@hermes.cam.ac.uk}
\affiliation{DAMTP, Centre for Mathematical Sciences, University of Cambridge, Wilberforce
Rd., Cambridge CB3 0WA, UK}
\keywords{Cosmology; Brans-Dicke; Wheeler-DeWitt; Group invariant transformations.}
\pacs{98.80.-k, 95.35.+d, 95.36.+x}

\begin{abstract}
We consider the application of group invariant transformations in order to
constrain a flat isotropic and homogeneous cosmological model, containing of a
Brans-Dicke scalar field and a perfect fluid with a constant equation of state
parameter $w$, where the latter is not interacting with the scalar field in
the gravitational action integral. The requirement that the Wheeler-DeWitt
equation be invariant under one-parameter point transformations provides us
with two families of power-law potentials for the Brans-Dicke field, in which
the powers are functions of the Brans-Dicke parameter $\omega_{BD}$ and the
parameter $w$. The existence of the Lie symmetry in the Wheeler-DeWitt
equation is equivalent to the existence of a conserved quantity in field
equations and with oscillatory terms in the wavefunction of the universe. This
enables us to solve the field equations. For a specific value of the conserved
quantity, we find a closed-form solution for the Hubble factor, which is
equivalent to a cosmological model in general relativity containing two
perfect fluids. This provides us with different models for specific values of
the parameters $\omega_{BD},$ and $w$. Finally, the results hold for the
specific case where the Brans-Dicke parameter $\omega_{BD}$ is zero, that is,
for the O'Hanlon massive dilaton theory, and consequently for $f\left(
R\right)  $ gravity in the metric formalism.

\end{abstract}
\maketitle

\section{Introduction}

The comprehensive analysis of various observational data (Cosmic Microwave
Background, Supernova Type Ia, large-scale structures, etc.) supports a
picture in which the universe is spatially flat with only about $30\%$ of its
total energy in the form of dark or luminous forms of matter. The nature of
the remaining $70\%$, residing in some unknown form of enigmatic 'dark energy'
remains a mystery even though it can be accurately described by a particular
type of anti-gravitating stress. Discovering the physics of this dark energy,
driving the recent accelerated expansion of the universe, is a key goal of
theoretical physics and cosmology. The intense debate among the cosmologists
and theoretical physics has opened up the possibility of various new
cosmological scenarios which offer different sources for the observed
acceleration and even a possibility to link it to the suspected era of
inflationary acceleration in the very early universe. Some of these scenarios
are based on the existence of new fields in nature, while others modify the
classical Einstein-Hilbert action for gravity
\cite{Ratra,Barrow93,Linder,Copeland,Overduin,Basil,lamel,cais,Bento,Brans,Buda,Sotiriou,Ferraro,Maartens,faraonibook,Ame10}%
.


In Friedmann-Lema\^{\i}tre-Robertson-Walker (FLRW) cosmologies containing
scalar fields, some analytical solutions without matter can be found in
\cite{jdb,ssf01,ssf02,ssf03,basilLukes,ssf04,ssf05}. If a matter component is
included in the dynamics then new solutions were also found in
\cite{ssf06,Barrow,Urena,sahni}. Furthermore, a new class of solutions has
been found from the application of group invariants, namely Lie/Noether
symmetries of the field equations \cite{basprd1,basprd2}. In fact the idea to
use Noether point symmetries in cosmology is not new and indeed there is a lot
of work in the literature (see \cite{Cap97}). In some previous papers we have
provided the Lie/Noether symmetries for various cosmological models, including
scalar fields \cite{basprd1}, $f(R)$ \cite{Tsafr}, $f(T)$ \cite{BASFT} and
scalar tensor theories \cite{AnST}. Recently, we have used this dynamical
symmetry approach in order to provide solutions to the WdW equation
\cite{prdB}.

In general for scalar-tensor theories, including the particular case of
Brans-Dicke gravity \cite{Brans}, various analytical solutions are available
in the literature \cite{Vernov,riazi,morg}. In Ref.\cite{clifton}, an exact
solution describes a Brans-Dicke scalar field which interacts with a perfect
fluid in the action integral. Moreover, some closed-form solutions for plane
symmetric spacetimes can be found in \cite{tiwari}, and some black-hole
solutions in Brans-Dicke gravity are given in \cite{bh01,bh02,bh03}. Using the
method of group invariants, some new exact solutions without matter source are
found in \cite{terzis1,Kam,capST}, and with dust in \cite{AnST}.

The purpose of this paper is to extend the method proposed in Ref.\cite{prdB}
to Brans-Dicke gravity when the perfect fluid does not interact with the
Brans-Dicke field in the action integral. The proposed selection rule
determines the potential that defines the Brans-Dicke field in order for the
Wheeler-DeWitt (WdW) equation to be invariant under a group of point
transformations. In \cite{prdB} that proposed method was applied in general
relativistic cosmology for a homogeneous scalar field and a perfect fluid. It
has been shown that when the WdW equation is invariant under one-parameter
point transformation then, the WdW equation can be solved by separation of
variables. The solution
provides oscillatory terms in the wavefunction and, at the same time, the
point transformations give Noetherian conservation laws for the classical
field equations. This latter property can be used to study the integrability
of the field equations and extract closed-form solutions.

The structure of the paper is as follows. In Section \ref{field} we present
the field equations in Brans-Dicke gravity. In Section \ref{sym} we apply Lie
point symmetries to the WdW equation and, in section \ref{solutions}, we
provide the invariant solution of the WdW equation. Also, in the same section,
we use the Hamilton-Jacobi theory to reduce the field equations to a pair of
first-order differential equations and under specific conditions (the
Noetherian conservation law vanishes) we obtain a closed-form solution for the
Hubble parameter in Brans-Dicke gravity. Then we check the performance of this
special Brans-Dicke model against the latest observational data. Finally, in
Section \ref{conc} we discuss our results and we draw our conclusions.

\section{Field equations in Brans-Dicke gravity}

\label{field}

In Brans-Dicke gravity the gravitational action in the Jordan frame and matter
that is not interacting with the Brans-Dicke field is defined by
\cite{faraonibook}%
\begin{equation}
S=\int dx^{4}\sqrt{-g}\left[  \frac{1}{2}\phi R-\frac{1}{2}\frac{\omega_{BD}%
}{\phi}g^{\mu\nu}\phi_{;\mu}\phi_{;\nu}-V\left(  \phi\right)  \right]  +\int
dx^{4}\sqrt{-g}L_{m}, \label{bd.01}%
\end{equation}
where $L_{m}$ is the Lagrangian of the matter, $\phi$ is the Brans-Dicke
scalar field, and $\omega_{BD}$ is the Brans-Dicke parameter.

Variations of $S$ (\ref{bd.01}) with respect to the metric tensor and the
field $\phi$ gives the modified Einstein field equations
\begin{equation}
\phi G_{\mu\nu}=\frac{\omega_{BD}}{\phi}\left(  \phi_{;\mu}\phi_{;\nu}%
-\frac{1}{2}g_{\mu\nu}g^{\kappa\lambda}\phi_{;\kappa}\phi_{;\lambda}\right)
-g_{\mu\nu}V\left(  \phi\right)  -\left(  g_{\mu\nu}g^{\kappa\lambda}%
\phi_{;\kappa\lambda}-\phi_{;\mu}\phi_{;\nu}\right)  +kT_{\mu\nu},
\label{bd.02}%
\end{equation}
and the Klein-Gordon equation%
\begin{equation}
g^{\mu\nu}\phi_{;\mu\nu}-\frac{1}{2\phi}g^{\mu\nu}\phi_{;\mu}\phi_{;\nu}%
+\frac{\phi}{2\omega_{BD}}\left(  R-2V_{,\phi}\right)  =0, \label{bd.04}%
\end{equation}
where $k\equiv8\pi G$. In (\ref{bd.02}) $T_{\mu\nu}$ is the energy-momentum
tensor of matter. Since we have assumed that the matter is not interacting
with the Brans-Dicke field, we have the conservation law $T_{~~~;\nu}^{\mu\nu
}=0.$

In the following we study the solution of these equations under the following assumptions.

a. Spacetime is spatially flat with FLRW metric line element%

\begin{equation}
ds^{2}=-dt^{2}+a^{2}\left(  t\right)  \left(  dx^{2}+dy^{2}+dz^{2}\right)  ,
\label{bd.05}%
\end{equation}
whose Ricci scalar is
\begin{equation}
R=6\left[  \frac{\ddot{a}}{a}+\left(  \frac{\dot{a}}{a}\right)  ^{2}\right]  .
\label{bd.06}%
\end{equation}
b. Matter is a perfect fluid for comoving observers $u_{\mu}=\delta_{\mu}^{0}%
$; that is, the energy-momentum tensor is%
\begin{equation}
T_{\mu\nu}=\left(  \rho_{m}+p_{m}\right)  u_{\mu}u_{\nu}+p_{m}g_{\mu\nu},
\label{bd.03}%
\end{equation}
where $\rho_{m},$ is the energy density of the matter and $p_{m}$ is the
isotropic pressure measured by the observers $u_{\mu}.$

c. The perfect fluid has a constant equation of state (EoS) parameter $w$,
i.e., $p_{m}=w\rho_{m}.$ For a barotropic fluid $w\in\left[  0,1\right]  $;
for $w=0$, $T_{\mu\nu}$, describes dust and for $w=1/3$, a radiation fluid. In
what follows we will extend the range of the EoS\ parameter to be $w\in\left(
-1,1\right)  $ and consider effects 'fluids'. Of course, the lower limit,
$w=-1$, corresponds to a cosmological constant which can always be absorbed in
the scalar-field potential.

d. We assume that the scalar field, $\phi$, possesses the same symmetries as
the spacetime, that is $\phi\left(  t,x,y,z\right)  \equiv\phi\left(
t\right)  $.

From the conservation law $T_{~~~;\nu}^{\mu\nu}=0$ we find that $\rho_{m}%
=\rho_{m0}a^{-3\left(  1+w\right)  }$, where $\rho_{m0}$ is the energy density
of today, and $\rho_{m0}=3\Omega_{m0}H_{0}^{2}$. Under these assumptions the
Lagrangian of the field equations becomes%
\begin{equation}
\mathcal{L}\left(  a,\dot{a},\phi,\dot{\phi}\right)  =-3a\phi\dot{a}%
^{2}-3a^{2}\dot{a}\dot{\phi}+\frac{1}{2}\frac{\omega_{BD}}{\phi}a^{3}\dot
{\phi}^{2}-a^{3}V\left(  \phi\right)  -k\rho_{m0}a^{-3w}, \label{bd.07}%
\end{equation}
and the first Brans-Dicke Friedmann equation is
\begin{equation}
3H^{2}=\frac{\omega_{BD}}{2}\left(  \frac{\dot{\phi}}{\phi}\right)  ^{2}%
+\frac{V\left(  \phi\right)  }{\phi}-3H\frac{\dot{\phi}}{\phi}+\frac{k}{\phi
}\rho_{m0}a^{-3\left(  1+w\right)  }, \label{bd.08}%
\end{equation}
where $H=\dot{a}/a$, is the Hubble function, and an overdot indicates
differentiation with respect to the comoving proper time coordinate $t$.

In general, the Lagrangian (\ref{bd.07}) defines the motion of a particle in a
two-dimensional space $\left\{  a,\phi\right\}  $ with effective potential%
\begin{equation}
V_{eff}=a^{3}V\left(  \phi\right)  +k\rho_{m0}a^{-3w}. \label{bd.09}%
\end{equation}
For $\omega_{BD}\neq-3/2$, from Lagrangian\footnote{In the limit in which
$\omega_{BD}=-3/2$, lagrangian (\ref{bd.07}) is denerate, i.e. $\left\vert
\frac{\partial^{2}L}{\partial\dot{x}^{i}\partial\dot{x}^{j}}\right\vert =0$,
for a discussion see \cite{sot7}.} (\ref{bd.07}) we define the momentum,
$p_{a}=\frac{\partial L}{\partial\dot{a}}$,~$p_{\phi}=\frac{\partial
L}{\partial\dot{\phi}}$, hence we can write the Hamiltonian
\begin{equation}
\mathcal{E}=\frac{1}{2\omega_{BD}+3}\left[  -\frac{\omega_{BD}}{2a\phi}%
p_{a}^{2}-\frac{3}{a^{2}}p_{a}p_{\phi}+\frac{3\phi}{a^{3}}p_{\phi}^{2}\right]
+a^{3}V\left(  \phi\right)  +k\rho_{m0}a^{-3w}. \label{bd.10}%
\end{equation}
From the first modified Friedmann equation (\ref{bd.08}) it follows that
$\mathcal{E}=0.$

Therefore, the second modified Friedmann equation and the Klein Gordon
equation are described by the following Hamiltonian system:%
\begin{equation}
\left(  2\omega_{BD}+3\right)  \dot{a}=-\frac{2\omega_{BD}}{3a\phi}p_{a}%
-\frac{2}{a^{2}}p_{\phi}, \label{bd.11}%
\end{equation}%
\begin{equation}
\left(  2\omega_{BD}+3\right)  \dot{\phi}=-\frac{2}{a^{2}}p_{a}+\frac{4\phi
}{a^{3}}p_{\phi}, \label{bd.12}%
\end{equation}%
\begin{equation}
\left(  2\omega_{BD}+3\right)  \dot{p}_{\phi}=-\frac{\omega_{BD}}{3a\phi^{2}%
}p_{a}^{2}-\frac{2}{a^{3}}p_{\phi}^{2}-\left(  2\omega_{BD}+3\right)
a^{3}V_{,\phi}, \label{bd.13}%
\end{equation}%
\begin{align}
\left(  2\omega_{BD}+3\right)  \dot{p}_{a}  &  =-\frac{\omega_{BD}}{3}%
\frac{p_{a}^{2}}{a^{2}\phi}-\frac{4}{a^{3}}p_{a}p_{\phi}+\frac{6\phi}{a^{4}%
}p_{\phi}^{2}+\nonumber\\
&  -3\left(  2\omega_{BD}+3\right)  \left(  a^{2}V\left(  \phi\right)
-k\rho_{m0}wa^{-1-3w}\right)  . \label{bd.14}%
\end{align}

Under normal quantization, i.e., $p_{i}\simeq i\frac{\partial}{\partial x^{i}%
},$ we can define the WdW equation~$W:=\mathcal{E}\left(  \Psi\right)  =0$,
that is \cite{wd1,HartleHaw},
\begin{align}
0  &  =\frac{1}{6\left(  2\omega_{BD}+3\right)  }\left[  \frac{\omega}{a\phi
}\Psi_{,aa}-\frac{6}{a^{2}}\Psi_{,a\phi}+\frac{6\phi}{a^{3}}\Psi_{,\phi\phi
}\right]  +\nonumber\\
&  +\frac{1}{6\left(  2\omega_{BD}+3\right)  }\left[  \frac{6}{a^{3}}%
\Psi_{,\phi}-\frac{\omega}{a^{2}\phi}\Psi_{,a}\right]  -\left[  a^{3}V\left(
\phi\right)  +k\rho_{m0}a^{-3w}\right]  \Psi, \label{bd.15}%
\end{align}
where $\Psi=\Psi\left(  a,\phi\right)  $ indicates the wavefunction of the
universe. Notice, that we use the following derivatives $\Psi_{,aa}%
=\partial^{2}\Psi/\partial a^{2}$, $\Psi_{\phi\phi}=\partial^{2}\Psi
/\partial\phi^{2}$ and $\Psi_{,a\phi}=\partial^{2}\Psi/\partial a\partial\phi
$. Recall that the dimension of the minisuperspace is two: that is, we do not
introduce the quantum correction term in order for the WdW equation to be
conformally invariant \cite{Wil,Hal}.

The WdW equation is defined by the conformally invariant operator
\begin{equation}
\hat{L}_{\gamma}=-\Delta_{\gamma}+\frac{n-2}{4\left(  n-1\right)  }R_{\gamma}
\label{bd.15a}%
\end{equation}
where $\Delta_{\gamma}$ is the Laplace operator with respect to the
minisuperspace $\gamma_{ij}$, and $R_{\gamma}$, is the Ricci scalar of
$\gamma_{ij}$. The importance of the operator $\hat{L}_{\gamma}$ is that under
a conformal transformation, $\bar{\gamma}_{ij}=e^{2\Omega\left(  x^{k}\right)
}\gamma_{ij},$ the scaling $\hat{L}_{\bar{\gamma}}\left(  \Psi\right)
=e^{-\frac{n+2}{2}\Omega\left(  x^{k}\right)  }\hat{L}_{\gamma}\left(
e^{\frac{n-2}{2}\Omega\left(  x^{k}\right)  }\Psi\right)  $ holds, where
$\Omega\left(  x^{k}\right)  $ is an arbitrary function. Moreover, using the
the kinetic term in the Lagrangian (\ref{bd.07}), we endow the minisuperspace
$\gamma_{ij}$ (the dimension is $n=2$), with line element:%
\begin{equation}
ds_{\left(  \gamma\right)  }^{2}=-6a\phi da^{2}-6a^{2}dad\phi+\frac
{\omega_{BD}}{\phi}a^{3}d\phi^{2},
\end{equation}
from which we conform that the associated Ricci scalar vanishes, i.e.
$R_{\gamma}=0.$

Hence, from (\ref{bd.15a}), it follows that $L_{\gamma}=-\Delta_{\gamma}$,
which is the Laplace operator, and under a conformal transformation we have
$L_{\bar{\gamma}}\left(  \Psi\right)  =e^{-2\Omega\left(  x^{k}\right)
}L_{\gamma}\left(  \Psi\right)  $. In order to solve (\ref{bd.15}), we need to
specify the scalar-field potential. This will be done by an ansatz. In the
literature there have been many forms for this potential depending on what one
wants to do. In the present work we adopt the geometric approach to dictate
the physics. The gain from this approach is that it is observer-free and no
conflicts arise between geometry and dynamics. Specifically, we require that
the potential $V(\phi)$ be such that the WdW equation (\ref{bd.15}) admits Lie
point symmetries.

\section{Group-invariant transformations for the WdW equation}

\label{sym}

For convenience, we provide below the basic definitions of Lie point
symmetries. Let $W=W\left(  x^{i},\Psi,\Psi_{,i},\Psi_{,ij}\right)  ~$be a
second-order differential equation, $x^{i}$ are the independent variables, and
$\Psi$ is the dependent variable, where $\Psi_{,i}=\frac{\partial\Psi
}{\partial x^{i}}$ and $\Psi_{,ij}=\frac{\partial^{2}\Psi}{\partial x^{i}
\partial x^{j}}$. The generator $X$ of the infinitesimal one-parameter point
transformation%
\begin{equation}
\bar{x}^{i}=x^{i}+\varepsilon\xi^{i}\left(  x^{i},\Psi\right)  ,~\bar{\Psi
}=\Psi+\varepsilon\eta\left(  x^{i},\Psi\right)  +O(\varepsilon^{2})
\label{bd.16}%
\end{equation}
is defined by
\begin{equation}
X=\frac{\partial\bar{x}^{i}}{\partial\varepsilon}\partial_{i}+\frac
{\partial\bar{\Psi}}{\partial\varepsilon}\partial_{\Psi}, \label{bd.17}%
\end{equation}
from which it follows that%
\begin{equation}
X=\xi^{i}\left(  x^{i},\Psi\right)  \partial_{i}+\eta\left(  x^{i}%
,\Psi\right)  \partial_{\eta}. \label{bd.18}%
\end{equation}

The differential equation, $W$,$~$is invariant under the action of the
one-parameter point transformation (\ref{bd.16}) if there exists a function
$\kappa$ such that \cite{Olver,IbragB}%
\begin{equation}
X^{\left[  2\right]  }W=\kappa W, \label{bd.20}%
\end{equation}
where $X^{\left[  2\right]  }$ is the second prolongation of $X$ in the jet
space $\left\{  x^{i},\Psi,\Psi_{,i},\Psi_{,ij}\right\}  $. When condition
(\ref{bd.20}) holds, we say that $X$ is a Lie point symmetry of $W$. Notice,
that the Lie point symmetries of a differential equation form a Lie algebra.

For differential equations which follow from a variational principle, i.e.
there exists a Lagrange function, the Lie point symmetries which transform the
action integral in such a way that the Euler-Lagrange equations are invariant
are called Noether point symmetries. The characteristic of Noether point
symmetries is that for each Noether symmetry, $X,$ there corresponds a
conservation law which is called a Noether integral \cite{EmmyN}. The Noether
point symmetries of a differential equation form a subalgebra of the Lie point
symmetries of that equation which is called the Noether algebra of the
differential equation.

\subsection{Lie point symmetries of the WdW equation}

The WdW equation (\ref{bd.15}) is a second-order partial differential equation
defined in the space of the independent variables $\left\{  x^{i}\right\}
\rightarrow\left\{  a,\phi\right\}  ,$ where $\Psi$ is the dependent variable.
Hence, the generator (\ref{bd.18}) of the infinitesimal point transformation
(\ref{bd.16}) has the following form,%
\begin{equation}
X=\xi^{a}\left(  a,\phi,\Psi\right)  \partial_{\alpha}+\xi^{\phi}\left(
a,\phi,\Psi\right)  \partial_{\phi}+\eta\left(  a,\phi,\Psi\right)
\partial_{\Psi}. \label{bd.21}%
\end{equation}

From condition (\ref{bd.20}), and for arbitrary $V\left(  \phi\right)  $, we
have the following Lie symmetries:%
\begin{equation}
X_{\Psi}=\Psi\partial_{\Psi},~X_{b}=b\left(  a,\phi\right)  \partial_{\Psi},
\label{bd.22}%
\end{equation}
where $b\left(  a,\phi\right)  $ is a solution of the original equation. The
vector field $X_{\Psi}$ is called a homogeneous symmetry, whereas $X_{b}$
corresponds to the infinite number of solutions. Both these Lie symmetries are
trivial symmetries in the sense that they cannot be used to reduce the
differential equation. However, they indicate that equation (\ref{bd.15}) is a
linear second-order partial differential equation.

In order that equation (\ref{bd.15}) admit non trivial Lie symmetries, we must
consider specific forms of the potential $V\left(  \phi\right)  .$ In
\cite{ijgmmp,prdB} it has been shown that the WdW equation (\ref{bd.15})
admits nontrivial Lie point symmetries if and only if the potential $V\left(
\phi\right)  $ is power law, with%
\begin{equation}
V\left(  \phi\right)  =V_{0}\phi^{\lambda}. \label{bd.23}%
\end{equation}
In our case the power $\lambda=\lambda\left(  \omega_{BD},w\right)  $ has the
following possible values: (a) $\lambda_{1}=\left(  1+w\right)  \left(
1-w\right)  ^{-1}$, and, (b) $2\lambda_{2}=\left(  \varpi-3\right)  \left(
w+1\right)  ,$ where~%
\begin{equation}
\varpi=\sqrt{6\omega_{BD}+9}. \label{bd.23a}%
\end{equation}

The Lie\ point symmetry vector which corresponds to $\lambda_{1}$
is\footnote{Recall that we have considered $w\in\left(  -1,1\right)  $. In the
limit where $w=1$, only the power law potential with $\lambda_{2}$ exists.}
\begin{equation}
X_{1}=a\partial_{a}+3\left(  w-1\right)  \phi\partial_{a}, \label{bd.24}%
\end{equation}
and to $\lambda_{2}$ is,%
\begin{equation}
X_{2}=A^{\mu_{1}}\phi^{\mu_{2}}\left(  a\partial_{a}+\frac{6\phi}{\varpi
-3}\partial_{\phi}\right)  , \label{bd.25}%
\end{equation}
where $\mu_{1}=\frac{3}{2\varpi}\left[  \varpi\left(  w-1\right)
-3w+1\right]  $ and $\mu_{2}=\frac{\varpi+3}{2}\mu_{1}$.

We remark that the vector fields $X_{1}$,~$X_{2}$, are conformal symmetries
for the minisuperspace $\gamma$ defined by the kinematic part of the
Lagrangian (\ref{bd.07}). Therefore, the minisuperspace selects the form of
the potential \cite{basprd1,prdB,ijgmmp}. A special case occurs when the
perfect fluid is radiation, that is, $w=1/3.$ In this case we find another
power-law potential of the form
\begin{equation}
V_{w=1/3}\left(  \phi\right)  =V_{0}\left[  \left(  V_{1}-\phi^{-\frac{1}%
{3}\varpi}\right)  ^{2}+V_{1}\left(  2-\phi^{\frac{1}{3}\varpi}\right)
^{2}-V_{1}\right]  . \label{bd.26}%
\end{equation}
For this potential the WdW equation admits the Lie point symmetry which is
given by the vector field, $Y=Y_{+}+Y_{-}~,~$where
\begin{equation}
Y_{\pm}=\phi^{-\frac{1}{2}\left(  1\pm\frac{\varpi}{3}\right)  }\left(
\partial_{a}+\frac{\varpi\pm3}{\omega_{BD}}\frac{\phi}{a}\partial_{\phi
}\right)  . \label{bd.27}%
\end{equation}
The importance of the existence of a nontrivial Lie point symmetry for
equation (\ref{bd.15}) is the existence of a coordinate system in which
equation (\ref{bd.15}) is independent of one of the independent variables, and
the solution which corresponds to the zero-order invariants has oscillatory
terms. Furthermore, a Noetherian conservation law for the field equations
(\ref{bd.10})-(\ref{bd.14}) means that they define an integrable Hamiltonian
system \cite{prdB}. In the next section we use the zero-order invariance on
the WdW equation (\ref{bd.15}) to reduce the field equations to two
first-order ordinary differential equations, which we solve.

\section{Analytical solutions}

\label{solutions}

Now we apply the zero-order invariants of the Lie point symmetries for the WdW
equation in order to reduce the equation and find the solution of the
wavefunction $\Psi(a,\phi)$. Moreover, by using the Hamilton-Jacobi equation
we reduce the dimension of the Hamiltonian system for the field equations. We
do that for the potential (\ref{bd.23}) with $\lambda=\lambda_{1}$.

\subsection{Invariant solution for the WdW equation}

In order to apply the zero-order invariants of a Lie point symmetry in
equation (\ref{bd.15}),\ we prefer to work with the normal coordinates of the
symmetry vector $X_{1}$. We apply the coordinate transformations%
\begin{equation}
a=\exp\left(  x\right)  ~,~\phi=y\exp\left[  3\left(  w-1\right)  x\right]
\label{bd.28a}%
\end{equation}
to (\ref{bd.15}) and the WdW equation takes the following form%
\begin{align}
0  &  =\left(  -\frac{\omega_{BD}}{3y}\Psi_{,xx}+2m_{1}\Psi_{,xy}-ym_{2}%
\Psi_{,yy}\right)  +\nonumber\\
&  ~~-m_{2}\Psi_{,y}-2\left(  \bar{V}_{0}y^{\frac{1+w}{1-w}}+\bar{\rho}%
_{m0}\right)  \Psi, \label{bd.29}%
\end{align}
where now we have $\Psi=\Psi\left(  x,y\right)  $, $m_{1}=3\omega_{BD}\left(
w-1\right)  -1,$ $m_{2}=3\omega_{BD}\left(  w-1\right)  ^{2}-2\left(
3w-2\right)  $ and $\left(  \bar{V}_{0},\bar{\rho}_{m0}\right)  =\left(
2\omega_{BD}+3\right)  \left(  V_{0},k\rho_{m0}\right)  $.

\begin{figure}[ptb]
\includegraphics[height=10cm]{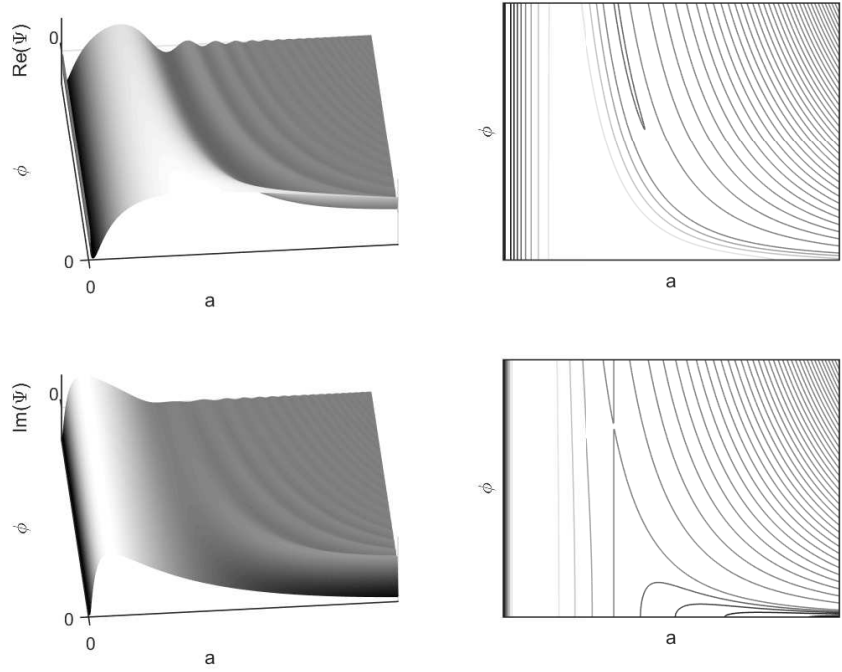}
\caption{\emph{Left panels:} The surface plot of the wavefunction
(\ref{bd.30}) for $\omega_{BD}=0$, $\bar{\rho}_{m0}=0$, $w=1/3$ and $\beta
=1i$, which corresponds to the quadratic potential $V\left(  \phi\right)
=V_{0}\phi^{2}$. Notice that we use $V_{0}=1$ units. The current dynamical
model effectively reduces to $f\left(  R\right)  \propto R^{2}$ gravity.
\emph{Right panels:} The contours of the wavefunction in the $(a,\phi)$
plane.}%
\label{wdw01}%
\end{figure}

In the new coordinates the Lie point symmetry vector $X_{1}$, takes the simple
form $X_{1}=\partial_{x}$. Since $X_{1}$ is a Lie symmetry of (\ref{bd.29})
this means that it transforms solutions to solutions; that is,
\begin{equation}
X_{1}\left(  \Psi\right)  =\beta\Psi, \label{bd.29a}%
\end{equation}
from where it follows that,\footnote{We take the same result when we apply the
zero-order invariants of the symmetry vector $Z=X_{1}-\beta X_{\Psi}$ in
(\ref{bd.29}).}
\begin{equation}
\Psi\left(  x,y\right)  =\sum\limits_{\beta}e^{-\beta x}\Phi\left(  y\right)
, \label{bd.30}%
\end{equation}
where the function, $\Phi\left(  y\right)  $, is given by the following
second-order ordinary differential equation:%

\begin{equation}
ym_{2}\Phi_{,yy}+\left(  2\beta m_{1}+m_{2}\right)  \Phi_{,y}+2\left(  \bar
{V}_{0}y^{\frac{1+w}{1-w}}+\bar{\rho}_{m0}+\frac{\omega_{BD}}{3y}\beta
^{2}\right)  \Phi=0. \label{bd.31}%
\end{equation}
In fig. \ref{wdw01} we present the wavefunction (\ref{bd.30}), in the case of
the O'Hanlon theory, i.e. with $\omega_{BD}=0;$ which can be seen as an
effective $f(R)$ gravity with $\phi=\frac{df}{dR}$ and $V\left(  \phi\right)
=\left(  \frac{df}{dR}R-f\right)  $ \cite{bcot}. Specifically,
in the left panels of fig. \ref{wdw01} we show the solutions of $Re\left(
\Psi\left(  a,\phi\right)  \right)  $ and $Im\left(  \Psi\left(
a,\phi\right)  \right)  $ respectively, for $\rho_{m0}=0$, and $w=1/3$. The
latter case is for the the radiation-dominated era. As we have discussed
above, the previously selected dynamical conditions imply $V\left(
\phi\right)  =V_{0}\phi^{2}$, hence we can easily show that the
corresponding\footnote{In $f\left(  R\right)  $-theory, in the metric
formalism the gravitational action integral is $S=\int dx^{4}\sqrt{-g}f\left(
R\right)  $,~where $R$ is the Ricci scalar of the underlying space with metric
$g_{ij}$.} $f\left(  R\right)  $-theory is $f\left(  R\right)  \propto R^{2}$.
We would like to remind the reader that the $R^{2}$ term provides a de-Sitter
behavior \cite{neban} which plays a critical role in the inflationary era. In
fact, if the gravitational Lagrangian is $f(R)$, there exist de Sitter
solutions of the theory with covariantly constant $R_{0}$ if it is a solution
of Barrow and Ottewill's condition $R_{0}f^{\prime}(R_{0})=2f(R_{0}),$
\cite{bott}, which is satisfied identically for the purely quadratic
Lagrangian. Notice, that in the right panels of fig. \ref{wdw01}, we plot the
contours of the wavefunction in the $(a,\phi)$ plane.

Now we focus on Eq.(\ref{bd.31}). In the special case of $m_{2}=0$, namely
$\omega_{BD}=\frac{2}{3}\frac{\left(  3w-2\right)  }{\left(  1-w\right)  ^{2}%
},$ the corresponding solution is
\begin{equation}
\Phi\left(  y\right)  =\Phi_{0}\exp\left[  -\bar{V}_{0}\frac{\left(
1-w\right)  }{2\beta m_{1}}y^{\frac{2}{1-w}}-\frac{\bar{\rho}_{m0}}{\beta
m_{1}}y-\frac{\beta}{6m_{1}}\ln y\right]  . \label{bd.32}%
\end{equation}

In the next section we continue with the classical solution of the field
equations. As we shall see, the constant $\beta$ is related to the value of
the Noetherian conservation law for the field equations, which corresponds to
the vector field $X_{1}$.

\subsection{Classical solution}

In the coordinate system (\ref{bd.28a}) the Hamiltonian of the field equations
(\ref{bd.10}) becomes%
\begin{equation}
\mathcal{E}=\frac{e^{-3wx}}{6}\left(  -\frac{\omega_{BD}}{y}p_{x}^{2}%
+6m_{1}p_{x}p_{y}-3ym_{2}p_{y}^{2}\right)  +e^{-3wx}\left(  \bar{V}%
_{0}y^{\frac{1+w}{1-w}}+\bar{\rho}_{m0}\right)  , \label{bd.33}%
\end{equation}
and the field equations are given by the following Hamiltonian system%
\begin{equation}
e^{3wx}\dot{x}=m_{1}p_{y}-\frac{\omega_{BD}}{3y}p_{x}, \label{bd.34}%
\end{equation}%
\begin{equation}
e^{3wx}\dot{y}=m_{1}p_{x}-m_{2}yp_{y}, \label{bd.35}%
\end{equation}%
\begin{equation}
e^{3wx}\dot{p}_{x}=3w\mathcal{E}, \label{bd.36}%
\end{equation}%
\begin{equation}
e^{3wx}y^{2}\dot{p}_{y}=\frac{1}{2}m_{2}y^{2}p_{y}^{2}+\frac{1}{6}\omega
_{BD}p_{x}^{2}-V_{0}\frac{\left(  1+w\right)  }{\left(  1-w\right)  }%
y^{\frac{2}{1-w}}. \label{bd.37}%
\end{equation}

From the first modified Friedmann equation, and equation (\ref{bd.36}), we
have that
\begin{equation}
p_{x}=I_{0}. \label{bd.38}%
\end{equation}
This is the Noetherian conservation law which corresponds to the symmetry
vector\footnote{As we can see we did not use the formulas of Noether theorems
to calculate the conservation law. However we derived it from the Hamilton
equations in the canonical coordinates of the vector field $X_{1}$.} $X_{1}$.
Comparing the last expression with (\ref{bd.29a}) we see that $\left\vert
\beta\right\vert \simeq\left\vert I_{0}\right\vert $.

From (\ref{bd.33}) we define the (null) Hamilton-Jacobi
equation:\footnote{Recall that the action $S\left(  x,y\right)  $ is related
to the momenta, $p_{x}=\frac{\partial S}{\partial x}$, and $p_{y}%
=\frac{\partial S}{\partial y}$.}%
\begin{equation}
\frac{1}{6}\left(  -\frac{\omega_{BD}}{y}\left(  \frac{\partial S}{\partial
x}\right)  ^{2}+6m_{1}\left(  \frac{\partial S}{\partial x}\right)  \left(
\frac{\partial S}{\partial y}\right)  -3ym_{2}\left(  \frac{\partial
S}{\partial y}\right)  ^{2}\right)  +\left(  \bar{V}_{0}y^{\frac{1+w}{1-w}%
}+\bar{\rho}_{m0}\right)  =0, \label{bd.39a}%
\end{equation}
where $S=S\left(  x,y\right)  $ is the action. Furthermore, from
(\ref{bd.38}), we have the constraint $\left(  \frac{\partial S}{\partial
x}\right)  =I_{0}$. Hence, from this and from (\ref{bd.39a}) we find,
\begin{align}
S\left(  x,y\right)   &  =I_{0}x+\frac{m_{1}}{m_{2}}I_{0}\ln\left(  y\right)
+\nonumber\\
&  +\int\sqrt{I_{0}^{2}y^{-2}\left[  \left(  \frac{m_{1}}{m_{2}}\right)
^{2}-\frac{\omega_{BD}\omega}{m_{2}}\right]  +2m_{2}^{-1}\left(  \bar{V}%
_{0}y^{\frac{2w}{1-w}}+\bar{\rho}_{m0}y^{-1}\right)  }dy, \label{bd.39}%
\end{align}
while for $m_{2}=0,$ the action $S\left(  x,y\right)  $ becomes
\begin{equation}
S\left(  x,y\right)  =I_{0}x+\frac{\omega_{BD}I_{0}}{6m_{1}}\ln\left(
y\right)  -\frac{1}{2}\left(  \bar{V}_{0}\left(  1-w\right)  y^{\frac{2}{1-w}%
}+\frac{\bar{\rho}_{m0}}{I_{0}m_{1}}y\right)  . \label{bd.40}%
\end{equation}

Hence, the field equations (\ref{bd.34})-(\ref{bd.37}) reduce to the following
two-dimensional system of first-order differential equations:%
\begin{equation}
e^{3wx}\dot{x}=m_{1}\left(  \frac{\partial S}{\partial y}\right)
-\frac{\omega_{BD}}{3y}\left(  \frac{\partial S}{\partial x}\right)  ,
\label{bd.40a}%
\end{equation}%
\begin{equation}
e^{3wx}\dot{y}=m_{1}\left(  \frac{\partial S}{\partial x}\right)
-m_{2}y\left(  \frac{\partial S}{\partial y}\right)  , \label{bd.40b}%
\end{equation}
where, $S=S\left(  x,y\right)  $, is given by (\ref{bd.39}) or (\ref{bd.40}).
In general, the solution of the system (\ref{bd.40a})-(\ref{bd.40b}), is not
given in closed form. Below, for the specific case where the Noetherian
conservation law vanishes, we can express the solution in terms of the scale factor.

\subsection{A special closed-form solution}

We consider now the case where \thinspace$I_{0}=0$, i.e. $p_{x}=0$ with
$m_{1}\neq0,$ From (\ref{bd.34}) and (\ref{bd.35}) we have $\frac{dy}%
{dx}=-\frac{m_{2}}{m_{1}}y,~$so $y=\phi_{0}e^{-\frac{m_{2}}{m_{1}}x}$, hence
with the use of (\ref{bd.28a}) for the field $\phi$, we have%
\begin{equation}
\phi\left(  a\right)  =\phi_{0}a^{M}~,~M=3\left(  w-1\right)  -\frac{m_{2}%
}{m_{1}} , \label{bd.41}%
\end{equation}
so $\dot{\phi}=\phi_{0}Ma^{M}H.$ Then, from (\ref{bd.08}), it follows that%
\begin{equation}
\left[  3\left(  1+M\right)  -\frac{\omega_{BD}M}{2}\right]  H^{2}%
=G_{eff}\left(  V_{0}^{\prime}a^{\frac{1+w}{1-w}M}+k\rho_{m0}a^{-3\left(
1+w\right)  }\right)  , \label{bd.42}%
\end{equation}
where $V_{0}^{\prime}=V_{0}\phi_{0}^{\frac{1+w}{1-w}},$ and $G_{eff}=\left[
\left(  3\left(  1+M\right)  -\frac{\omega_{BD}M}{2}\right)  \phi\right]
^{-1}$ is the effective gravitational constant. We see that $G_{eff}\left(
a\rightarrow1\right)  =\left[  \left(  3\left(  1+M\right)  -\frac{\omega
_{BD}M}{2}\right)  \phi_{0}\right]  ^{-1}$. Therefore, we can say that when
$I_{0}=0$, the scalar field behaves like an effective fluid with constant EoS
parameter. A special solution of the form, $\phi=\phi_{0}a^{\phi_{1}}$, where
$\phi_{1}$ is a constant, has been found in \cite{clifton}, where $V_{0}=0$
and the perfect fluid is interacting with the scalar field in the action
integral. A similar result has been found in \cite{prdB}
. In that paper we applied the same geometric selection rule for the scalar
field, but in general relativity ($G_{eff}=const$)
as a special solution of the field equations\footnote{For a different
derivation of the same result see \cite{Barrow,Urena,sahni}.}.

By replacing $G_{eff}\left(  a\right)  $ in (\ref{bd.42}), we can define the
Hubble function as follows
\begin{equation}
H\left(  a\right)  ^{2} =H_{0}^{2} \left(  \Omega_{\phi0}a^{q_{1}}+\Omega
_{m0}a^{q_{2}} \right)  , \label{bd.43}%
\end{equation}
where spatial flatness requires that $\Omega_{\phi0}+\Omega_{m0}=1,$ since
$E\left(  a\rightarrow1\right)  =1$. Furthermore, the new constants$~q_{1}%
,q_{2}$ are,
\begin{equation}
q_{1}=\frac{2M\left(  w,\omega_{BD}\right)  w}{1-w}, \label{bd.44}%
\end{equation}%
\begin{equation}
q_{2}=-\left[  M\left(  w,\omega_{BD}\right)  +3\left(  1+w\right)  \right]  .
\label{bd.45}%
\end{equation}
Hence, the system (\ref{bd.44})-(\ref{bd.45}) can provide us with a Hubble
function for different models of two fluids (\ref{bd.43}). We study some
special cases:

Case (A): Cosmological constant with dust. This means that $\left(
q_{1},q_{2}\right)  =\left(  0,-3\right)  $ or~$\left(  q_{1},q_{2}\right)
=\left(  -3,0\right)  $, from which we have $\left(  w,\omega_{BD}\right)
=\left(  0,\frac{1}{6}\right)  $ or $\left(  w,\omega_{BD}\right)
\simeq\left(  0.28,-0.77\right)  $.

Case (B): Dust with radiation. This requires, $\left(  q_{1},q_{2}\right)
=\left(  -3,-4\right)  $ or $\left(  q_{1},q_{2}\right)  =\left(
-4,-3\right)  ,$ hence $\left(  w,\omega_{BD}\right)  \simeq\left(
0.63,0\right)  $ or $\left(  w,\omega_{BD}\right)  \simeq\left(
0.55,-1\right)  $.

Case (C): Cosmological constant with radiation fluid This requires $\left(
q_{1},q_{2}\right)  =\left(  0,-4\right)  $ or~$\left(  q_{1},q_{2}\right)
=\left(  -4,0\right)  $, hence $\left(  w,\omega_{BD}\right)  =\left(
0,0\right)  ,~\left(  w,\omega_{BD}\right)  =\left(  \frac{1}{3},0\right)  $
or,$~\left(  w,\omega_{BD}\right)  =\left(  \frac{1}{3},-\frac{3}{4}\right)  $.

Case (D): In the case of $\left(  q_{1},q_{2}\right)  =\left(  0,0\right)  $,
which implies $\left(  w,\omega_{BD}\right)  =\left(  -1,\frac{1}{6}\right)  $
or $\left(  w,\omega_{BD}\right)  =\left(  0,-\frac{4}{3}\right)  $, from
(\ref{bd.43}) we have a de Sitter solution.

It is interesting that when we assume a radiation fluid in (\ref{bd.43}) we
have a solution of the system (\ref{bd.44})-(\ref{bd.45}) in which
$\omega_{BD}=0$; however, this result is expected since when $\omega_{BD}=0$,
the action (\ref{bd.01}) reduces to O'Hanlon's massive dilaton gravity
\cite{Hanlon}, and consequently to $f\left(  R\right)  $-gravity in the metric
formalism, which provides a radiation term \cite{Ame10}.

Before we close this section, we should add that for the power-law potential
(\ref{bd.23}) with $\lambda=\lambda_{2}$, one may use the same method to
construct the solution of the field equations. We shall not repeat the
calculations but we simply say that in this case the canonical coordinate
transformation $\left\{  a,\phi\right\}  \rightarrow\left\{  z,r\right\}  $ is
given by the following expression:%
\begin{equation}
z=\frac{3-\varpi}{6\mu_{2}+\mu_{1}\left(  \varpi-3\right)  }a^{-\mu_{1}}%
\phi^{-\mu_{2}}~,~r=\phi a^{-\frac{6}{\varpi-3}}.
\end{equation}

\subsection{Observational Constraints}

Now we focus on the Hubble parameter (\ref{bd.43}) in which we have imposed
$w=0$. This means that the perfect fluid in the gravitation action
(\ref{bd.01}) is dust. Therefore, from (\ref{bd.44}) and (\ref{bd.45}) we have
that $q_{2}=-\frac{3\omega_{BD}+4} {3\omega_{BD}+1}$ and $q_{1}=0$. Using the
above conditions the Hubble parameter becomes
\begin{equation}
H\left(  a\right)  ^{2}=H_{0}^{2}\left[  \left(  1-\Omega_{m0}\right)
+\Omega_{m0}a^{-\frac{3\omega_{BD}+4}{3\omega_{BD}+1}}\right]  . \label{bd.46}%
\end{equation}
We mention that from the second term of Eq.(\ref{bd.46}) one may define an
effective equation of state parameter, namely
\[
w_{m}^{\mathrm{(eff)}}=\frac{1}{3}\frac{1-6\omega_{BD}}{\left(  3\omega
_{BD}+1\right)  }.
\]
Obviously, for $\omega_{BD}=1/6$ (or $w_{m}^{\mathrm{(eff)}}=0$) the above
Hubble parameter reduces to that of the concordance $\Lambda$CDM model.


In order to constrain the Brans-Dicke parameter we perform a joint likelihood
analysis using the Type Ia supernova data set of Union 2.1 \cite{Suzuki}, and
the BAO data \cite{Percival,BlakeC}. Notice, that for the Hubble constant we
utilize $H_{0}=69.6$km/s/Mpc \cite{BennetH0}. Hence, the overall likelihood
function is defined as follows%
\begin{equation}
\mathcal{L}\left(  \Omega_{m0},\omega_{BD}\right)  \mathcal{=L}_{SNIa}%
\mathcal{\times L}_{BAO}%
\end{equation}
where $\mathcal{L}_{A}\varpropto e^{-\chi_{A}^{2}/2}~$ which means that the
total $\chi^{2}$ is written as
\begin{equation}
\chi^{2}=\chi_{SNIa}^{2}+\chi_{BAO}^{2}.
\end{equation}
Lastly, in order to test the performance of the cosmological models against
the data we use the Akaike information criterion AIC$=\mathrm{min}(\chi
^{2})+2n_{fit}$, where $n_{fit}$ is the number of free parameter
\cite{Akaike1974}.

In the case of SNIa the corresponding chi-square parameter is given
by\footnote{For the SNIa test we have applied the diagonal covariant matrix
without the systematic errors.}%
\begin{equation}
\chi_{SNIa}^{2}=\sum\limits_{i=1}^{N_{SNIa}}\left(  \frac{\mu_{obs}\left(
z_{i}\right)  -\mu_{th}\left(  z_{i};\Omega_{m0},\omega_{BD}\right)  }%
{\sigma_{i}}\right)  ^{2}%
\end{equation}
where $N_{SNIa}=580$, $z_{i}\in\lbrack0.015,1.414]$ is the observed redshift,
$\mu_{obs}$ is the observed distance modulus and $\mu_{th}=5logD_{L}+25$ with
$D_{L}$ denoting the luminosity distance. Furthermore, for BAOs the
corresponding chi-square parameter has the following form
\begin{equation}
\chi_{BAO}^{2}=\sum\limits_{i=1}^{N_{BAO}}\left(  \sum\limits_{j=1}^{N_{BAO}%
}\left[  d_{obs}\left(  z_{i}\right)  -d_{th}\left(  z_{i};\Omega_{m0}%
,\omega_{BD}\right)  \right]  C_{ij}^{-1}\left[  d_{obs}\left(  z_{j}\right)
-d_{th}\left(  z_{j};\Omega_{m0},\omega_{BD}\right)  \right]  \right)
\end{equation}
where $N_{BAO}=6$ and $C_{ij}^{-1}$ is the inverse of the covariant matrix in
terms of $d_{z}=\frac{l_{BAO}}{D_{V}\left(  z\right)  }$ \cite{BasilNess}.
Notice, that the quantity $l_{BAO}\left(  z_{drag}\right)  $ is the BAO scale
at the drag redshift and $D_{V}\left(  z\right)  $ is the volume distance
\cite{BlakeC}.

\begin{figure}[ptb]
\includegraphics[height=7cm]{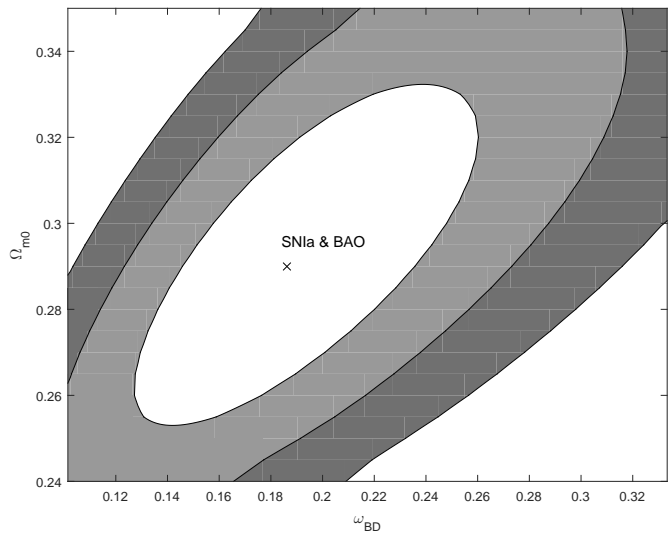}
\caption{The combined (SNIa+BAOs) likelihood contours $1\sigma\left(
\Delta\chi^{2}=2.3\right)  $, $2\sigma\left(  \Delta\chi^{2}=6.18\right)  $,
and $3\sigma\left(  \Delta\chi^{2}=11.83\right)  $, in the $\left(
\Omega_{m0},\omega_{DE}\right)  $ plane. In this analysis we use the Hubble
parameter of Eq. (\ref{bd.46}). The cross corresponds to best fit solution.}%
\label{confi}%
\end{figure}

In table \ref{table1}, we present the results of the current statistical
analysis while in figure (\ref{confi}) we provide the $1\sigma$, $2\sigma$,
and $3\sigma$ combined likelihood contours for the Brans-Dicke model [see
Eq.(\ref{bd.46})].
In particular, we find the following results:

\begin{itemize}
\item for the Brans-Dicke model: $\Omega_{m0}=0.29^{+0.032}_{-0.025}$,
$\omega_{BD}=0.19^{+0.075}_{-0.059}$, $\mathrm{min}(\chi^{2})\simeq564.29$,
$n_{fit}=2$ and AIC$\simeq568.29$.

\item for the $\Lambda$CDM model: $\Omega_{m0}=0.28^{+0.025}_{-0.024}$,
$\mathrm{min}(\chi_{\Lambda}^{2}) \simeq564.51$, $n_{fit}=1$ and
AIC$_{\Lambda}$$\simeq566.51$.
\end{itemize}

Since, $\Delta\mathrm{AIC}=|\mathrm{AIC}-\mathrm{AIC}_{\Lambda}| \le2$ we
conclude that the current cosmological models fit equally well the
observational data.

\begin{table}[tbp] \centering
\caption{The overall statistical results
(using SNIa+BAO) for the $\Lambda$CDM and Brans-Dicke models
respectively. Notice, that in our analyis we use Eq.(\ref{bd.46}).
In the last three colums we present the number of
free parameters and the goodness-of-fit statistics.}%
\begin{tabular}
[c]{ccccccccc}\hline\hline
Model & $\Omega_{m0}$ & $\Omega_{\Lambda}$ & $w_{\Lambda}~$ & $w_{m}$ &  &
$n_{fit}$ & $\min\left(  \chi^{2}\right)  $ & $\mathrm{AIC}$\\\hline
$\Lambda$CDM & $0.28_{-0.024}^{+0.025}$ & $0.72_{-0.024}^{+0.025}$ & $-1$ &
$0$ &  & $1$ & \thinspace$564.51$ & \thinspace$566.51$\\
&  &  &  &  &  &  &  & \\
& $\Omega_{m0}$ & $\Omega_{\phi0}$ & $w_{\phi}$ & $w_{m}^{(\mathrm{eff)}}$ &
$\omega_{BD}$ & $n_{fit}$ & $\min(\chi^{2})$ & $\mathrm{AIC}$\\\hline
Brans-Dicke & $0.29_{-0.025}^{+0.032}$ & $0.71_{-0.025}^{+0.032}$ & $-1$ &
$-0.03_{-0.072}^{+0.091}$ & $0.19_{-0.059}^{+0.075}$ & $2$ & $564.29$ &
$568.29$\\\hline\hline
\end{tabular}
\label{table1}%
\end{table}%


\section{Conclusions}

\label{conc}

In this paper, we have extended our earlier analysis, which was introduced in
\cite{prdB}, for the case of Brans-Dicke gravity with a perfect fluid in which
the fluid EoS parameter is constant and the underlying geometry is that of a
spatially flat FLRW universe. In particular, in order to select the functional
form of the scalar field potential, $V\left(  \phi\right)  ,$ in the
gravitational action integral (\ref{bd.01}), we have used the well known
criterion, namely the existence of group invariant transformations for the WdW
equation \cite{prdB}. The existence of a Lie point symmetry vector for the WdW
equation is related to the existence of oscillatory terms in the solution of
the wavefunction $\Psi$, and to Noetherian conservation laws for the field
equations. The latter can be used to find analytical solutions.

For our model, in which the perfect fluid is not interacting with the scalar
field, we found two families of solutions with power-law potential $V\left(
\phi\right)  =V_{0}\phi^{\lambda}$, where the constant $\lambda$ depends on
the EoS parameter, $w$, of the perfect fluid and the Brans-Dicke parameter
$\omega_{BD}$. \ The results hold for the case where the parameter
$\omega_{BD}$ vanishes, i.e., action (\ref{bd.01}) is that of O'Hanlon theory
which is equivalent with $f\left(  R\right)  $-gravity in the metric
formalism. As a special case, a third family of power potentials is found when
the perfect fluid is radiation, i.e. $w=1/3$.

By applying the zero-order invariants of the corresponding Lie symmetry in the
WdW equation, we were able to solve the WdW equation and find the oscillatory
behavior of the wave function. Furthermore, with the use of the
Hamilton-Jacobi theory we reduced the Hamiltonian system, which defines the
classical field equations, to a system of two first-order differential
equations. That is, the field equations for that form of the potential,
$V\left(  \phi\right)  $, form a Liouville integrable dynamical system.

As a special case for one of the integrable models, we found a closed-form
solution for the Hubble function from which we saw that the Brans-Dicke field
follows power-law behaviour with respect to the metric scale factor, that is,
$\phi=\phi_{0}a^{M}$. In this context, the Hubble parameter is written as
$H(a)=H_{0}\sqrt{\Omega_{\phi0}a^{q_{1}}+\Omega_{m0}a^{q_{2}}}$, where the
parameters $q_{1,2}$ are given in terms of $(w,\omega_{BD})$. Obviously, one
may recover a similar Hubble parameter to this within the framework of general
relativistic cosmology by using two different perfect fluids with constant
equation of state parameters. This implies that the current Brans-Dicke
gravity model is cosmologically equivalent to that of general relativity as
far as the cosmic expansion is concerned. Therefore, in order to distinguish
Brans-Dicke gravity from GR we need to extend the analysis to the perturbation
level. It is well known that modified gravity affects, via the effective
gravitational constant the growth rate of linear matter perturbations. As an
example, in the case of $f(R)$ gravity models the quantity $G_{eff}$ is given
in terms of the scale factor and of the wave-number (see \cite{BasilNess}%
,\cite{GANN} and references therein). In a forthcoming paper, we attempt to
investigate the impact of the Brans-Dicke gravitational parameter
$G_{eff}=\left[  \left(  3\left(  1+M\right)  -\frac{\omega_{BD}M}{2}\right)
\phi\right]  ^{-1}$ on the matter perturbations. Lastly, performing a joint
statistical analysis involving the recent SNIa and BAO data, we place
constraints on the main cosmological parameters of the Brans-Dicke model.

It has also been shown that there exists a relation between the value of the
conserved quantity,~$I_{0}$, for the classical field equations and the
\textquotedblleft frequency\textquotedblright,~$\beta$, in the WdW equation.
Hartle proposed that strong peaks in the wavefunction lead to a classical
observable universe \cite{hartle}. Since the WdW equation is a linear
second-order PDE, the general invariant solution is the sum on all possible
values of $\beta$. However, one may relate the \textquotedblleft
frequency\textquotedblright\ of the strong peaks of the wavefunction to the
value of the conserved quantity, $I_{0}$, (\ref{bd.38}). The latter is
included in the solution of the field equations, i.e. $a\left(  t\right)  $,
and consequently in the Hubble function $H\left(  a\right)  $. In any case,
the existence of the conservation law indicates a strong relation between the
classical and the quantum solutions and information can be transferred from
among the two systems. However, the physical observable quantities which
correspond to the conservation laws of the field equations are still unknown.

The Brans-Dicke action is defined in the Jordan frame and is conformally
equivalent to the minimally coupled scalar field in the Einstein frame. The
solutions which we have found describe a perfect fluid which is not
interacting with the scalar field in the Jordan frame. However, the solutions
can be transformed into the Einstein frame and will hold for a model in which
there exists an interaction between the perfect fluid and the scalar field of
a particular form. By definition, the two models share the same solution of
the WdW equation.

In a forthcoming work we will study the cosmological evolution of these
integrable models in the Einstein frame.


\begin{acknowledgments}
AP thanks the University of Athens for the hospitality provided there when
this work done. AP is supported by FONDECYT postdoctoral grant no. 3160121. SB
acknowledges support by the Research Center for Astronomy of the Academy of
Athens in the context of the program \textit{\textquotedblright Tracing the
Cosmic Acceleration\textquotedblright}. JDB is supported by the STFC.
\end{acknowledgments}


\begin{thebibliography}{99}                                                                                               %


\bibitem {Ratra}B. Ratra and P.J.E. Peebles, Phys. Rev. D \textbf{37,} 3406 (1988)

\bibitem {Barrow93}J.D. Barrow and P. Saich, Class. Quant. Grav. \textbf{10,}
279 (1993)

\bibitem {Linder}E.V. Linder, Phys. Rev. Lett. \textbf{90,} 091301 (2003)

\bibitem {Copeland}E.J. Copeland, M. Sami and S. Tsujikawa, Int. J. Mod. Phys.
D \textbf{15,} 1753 (2006)

\bibitem {Overduin}J.M. Overduin and F.I. Cooperstock, Phys. Rev. D
\textbf{58,} 043506 (1998)

\bibitem {Basil}S. Basilakos, M. Plionis and J. Sola, Phys. Rev. D
\textbf{80,} 083511 (2009)

\bibitem {lamel}L. Amendola, M. Baldi and C. Wetterich, Phys. Rev. D
\textbf{78, }023015 (2008)

\bibitem {cais}Y.F. Cai, E.N. Saridakis, M.R. Setare and J.Q. Xia, Phys. Rept.
\textbf{493,} 1 (2010)

\bibitem {Bento}M.C. Bento, O. Bertolami and A.A. Sen, Phys. Rev. D
\textbf{66,} 043507 (2002)

\bibitem {Brans}C. Brans and R.H. Dicke, Phys. Rev. \textbf{124,} 195 (1961)

\bibitem {Buda}H.A. Buchdahl, Mon. Not. Roy. Astron. Soc. \textbf{150,} 1 (1970)

\bibitem {Sotiriou}T.P. Sotiriou and V. Faraoni Rev. Mod. Phys. \textbf{82,}
451 (2010)

\bibitem {Ferraro}R. Ferraro and F. Fiorini, Phys. Rev. D \textbf{75,} 084031 (2007)

\bibitem {Maartens}R. Maartens, Living Rev. Rel. \textbf{7,} 7 (2004)

\bibitem {faraonibook}V. Faraoni, \textit{Cosmology in Scalar-Tensor Gravity},
Fundamental Theories of Physics vol. 139, (Kluwer Academic Press: Netherlands, 2004)

\bibitem {Ame10}L. Amendola and S. Tsujikawa, \textit{Dark Energy Theory and
Observations}, (Cambridge University Press: Cambridge, 2010)

\bibitem {jdb}J.D. Barrow, Phys. Lett. B \textbf{235}, 40 (1990)

\bibitem {ssf01}A.G. Muslinov, Class. Quantum Gravit. \textbf{13,} 3229 (1996)

\bibitem {ssf02}G.F.R. Ellis and M.S. Madsen, Class.\ Quantum Gravit.\textbf{
8,} 667 (1991)

\bibitem {ssf03}J.G. Russo, Phys. Lett. B \textbf{600,} 185 (2004)

\bibitem {basilLukes}S. Basilakos and G. Lukes-Gerakopoulos, Phys. Rev. D
\textbf{78,} 083509 (2008)

\bibitem {ssf04}J.J. Halliwell, Phys. Lett. B \textbf{185,} 341 (1987)

\bibitem {ssf05}R. Easter, Class. Quantum Gravit. 10 2203 (1993)

\bibitem {ssf06}L.P. Chimento and A.S. Jakubi, Int. J. Mod. Phys. D
\textbf{5,} 71 (1996)

\bibitem {Barrow}C. Rubano and J. D. Barrow, Phys. Rev. D. \textbf{64,} 127301 (2001)

\bibitem {Urena}L. A. Urena-Lopez, T. Matos, Phys. Rev. D \textbf{62,} 081302 (2000)

\bibitem {sahni}V. Sahni and A. Starobinsky, Int. J. Mod. Phys. D \textbf{9,}
373 (2000)

\bibitem {basprd1}S. Basilakos, M. Tsamparlis and A. Paliathanasis, Phys. Rev.
D \textbf{83,} 103512 (2011)

\bibitem {basprd2}A. Paliathanasis, M. Tsamparlis and S. Basilakos, Phys. Rev.
D \textbf{90,} 103524 (2014)

\bibitem {Cap97}S. Capozziello, R. de Ritis and A. A. Marino, Class. Quantum
Gravity, \textbf{14}, 3259 (1997); C. Rubano, and P. Scudellaro, Gen. Relat.
Grav. \textbf{34}, 307, (2002); A.K. Sanyal, B.Modak, C. Rubano and E.
Piedipalumbo, Gen. Relat. Grav. \textbf{37}, 407, (2005); M. Szydlowski et
al., Gen. Rel. Grav., \textbf{38}, 795, (2006); S. Capozziello, A. Stabile,
and A. Troisi, Class. Quant. Grav., \textbf{ 24}, 2153, (2007); A. Bonanno, G.
Esposito, C. Rubano and P. Scudellaro, Gen. Rel. Grav., \textbf{39}, 189,
(2007); S. Capozziello, E. Piedipalumbo, C. Rubano, and, P. Scudellaro, Phys.
Rev. D., \textbf{80}, 104030, (2009); B. Vakili, Phys. Lett. B., \textbf{664},
16, (2008); Yi Zhang, Yun-gui Gong and Zong-Hong Zhu, Phys. Lett. B.,
\textbf{688}, 13, (2010); Hao Wei, Xiao-Jiao Guo and Long-Fei Wang, Phys.
Lett. B., \textbf{707}, 298, (2010)

\bibitem {Tsafr}A. Paliathanasis, M. Tsamparlis and S. Basilakos, Phys. Rev. D
\textbf{84}, 123514, (2011).

\bibitem {BASFT}S. Basilakos, S. Capozziello, M. De Laurentis, A.
Paliathanasis, M. Tsamparlis, Phys. Rev. D., \textbf{88}, 103526 (2013)

\bibitem {AnST}A. Paliathanasis, M. Tsamparlis, S. Basilakos and S.
Capozziello, Phys. Rev. D \textbf{89,} 063532 (2014)

\bibitem {prdB}A. Paliathanasis, M. Tsamparlis, S. Basilakos and J.D. Barrow,
Phys. Rev. D \textbf{91,} 123535 (2015)

\bibitem {Vernov}A. Yu Kamenshchik, E.O. Pozdeeva, A. Tronconi, G. Venturi and
S.\ Yu Vernov, Class. Quantum Gravit. \textbf{31,} 105003 (2014)

\bibitem {riazi}E. Ahmadi-Azar and N. Riazi, Astr. Sp. Sci. \textbf{226,} 1 (1995)

\bibitem {morg}R.E. Morganstern, Phys. Rev. D \textbf{4,} 946 (1971)

\bibitem {tiwari}R.N. Tiwari and B.K. Nayak, J. Phys. A: Math. Gen.
\textbf{9,} 369 (1976)

\bibitem {bh01}H. Kim, Phys. Rev. D \textbf{60,} 024001 (1999)

\bibitem {bh02}N. Riazi and H.R. Askari, Mon. Not. R. Astron. Soc.
\textbf{261,} 229 (1993)

\bibitem {bh03}T. Clifton, J.D. Barrow and R.J. Scherrer Phys. Rev. D
\textbf{71}, 123526 (2005)

\bibitem {Kuku}Y. Kucukakca, U. Camci and I. Semiz, Gen. Rel. Gravit.,
\textbf{44,} 1893 (2012)

\bibitem {terzis1}P.A. Terzis, N. Dimakis and T. Christodoulakis, Phys.\ Rev.
D. \textbf{90,} 123543 (2014)

\bibitem {Kam}S. Kamilya and B. Modak, Gen. Rel. Gravit. \textbf{36,} 674 (2004)

\bibitem {capST}S. Capozziello, S. Nesseris and L. Perivolaropoulos, JCAP
\textbf{2012} 12(9) (2012)

\bibitem {clifton}T. Clifton and J.D. Barrow, Phys. Rev. D \textbf{73,} 104022 (2006)

\bibitem {sot7}T.P. Sotiriou, Gravity and Scalar fields, \textit{Proceedings
of the 7th Aegean Summer School: Beyond Einstein's theory of gravity},
Modifications of Einstein's Theory of Gravity at Large Distances, Paros,
Greece, ed. by E. Papantonopoulos, Lect.Notes Phys. \textbf{892} (2015)

\bibitem {wd1}B.S. DeWitt, Phys. Rev. \textbf{160} 1113 (1967)

\bibitem {HartleHaw}J.B. Hartle and S.W. Hawking, Phys. Rev. D \textbf{28}
2960 (1983)

\bibitem {Wil}D. Wiltshire, An introduction to quantum cosmology (2001)
(arXiv: gr-qc/0101003);

\bibitem {Hal}J.J. Halliwell, Introductory lectures on quantum cosmology
(2009) (arXiv: 0909.2566 [gr-qc])

\bibitem {Olver}P.J. Oliver, \textit{Applications of Lie Groups to
Differential Equations}, Graduate Texts in Mathematics, Volume \textbf{107},
(Springer-Verlag: New York, 1993)

\bibitem {IbragB}N.H. Ibragimov, \textit{Transformation Groups Applied to
Mathematical Physics Mathematics and Its Applications}, Soviet Series, (D
Reidel Publishers: Dordrecht, 1985)

\bibitem {EmmyN}E. Noether, Invariante Variationsprobleme, Nachr. v.d. Ges. d.
Wiss. zu G\"{o}ttingen, \textbf{235} (1918)

\bibitem {ijgmmp}A. Paliathanasis and M. Tsamparlis, Int. J. Geom. Meth. Mod.
Phys. \textbf{11} 1450037 (2014)

\bibitem {Hanlon}J. O'Hanlon, Phys. Rev. Lett. \textbf{29} 137 (1972)

\bibitem {bcot}J.D. Barrow and S. Cotsakis, Phys. Lett. B \textbf{214}, 515 (1988)

\bibitem {neban}A. Paliathanasis, J. Phys.: Conf. Ser. \textbf{453} 012009 (2013)

\bibitem {bott}J.D. Barrow and A.C. Ottewill, J. Phys. A \textbf{16}, 2757 (1983)

\bibitem {Suzuki}N. Suzuki et.al, Astrophys. J. 746 85 (2012)

\bibitem {Percival}W.J. Percival et al, Mon. Not. R. Astron. Soc.,
\textbf{401} 2148 (2010)

\bibitem {BlakeC}C. Blake et al., Mon. Not. R. Astron. Soc. \textbf{418} 1707 (2011)

\bibitem {BennetH0}C.L. Bennet, D. Larson, J.L. Weiland and G. Hinshaw, Ap. J.
\textbf{794 }135 (2014)

\bibitem {BasilNess}S. Basilakos, S. Nesseris and L. Perivolaropoulos, Phys.
Rev. D. \textbf{87}~123529 (2013)

\bibitem {Akaike1974}H. Akaike, IEEE Transactions of Automatic Control,
\textbf{19}, 716 (1974); N. Sugiura, Communications in Statistics A, Theory
and Methods, \textbf{7}, 13 (1978)

\bibitem {hartle}J.B. Hartle, \textit{Gravitation in Astrophysics: Carg\`{e}se
1986}, Proceedings of a Nato Advanced Study Institute on Gravitation in
Astrophysics, Carg\`{e}se, France, Ed. by B. Carter and J.B. Hartle, (Plenum:
New York, 1986)

\bibitem {GANN}R. Gannouji, B. Moraes and D. Polarski, JCAP, \textbf{62}, 034
(2009); S. Tsujikawa, R. Gannouji, B. Moraes and D. Polarski, Phys. Rev. D.,
\textbf{80}, 084044 (2009)








\end{thebibliography}
\end{document}